\documentclass[twocolumn,superscriptaddress,amsmath,amssymb,aps,pra,floatfix]{revtex4-1}
\usepackage[utf8]{inputenc}
\usepackage[english]{babel}
\usepackage{wrapfig}
\usepackage{balance}
\usepackage{flushend}
\usepackage{float}
\usepackage{amssymb}
\usepackage{amsfonts}
\usepackage{amsmath}
\usepackage{cases}
\usepackage{stmaryrd}\usepackage{tipa}
\usepackage[dvips]{graphicx}
\usepackage{dcolumn}
\usepackage{bbold}
\usepackage{graphicx}
\usepackage{graphicx}
\usepackage{subfigure}
\usepackage{dcolumn}
\usepackage{bm}
\usepackage{color}
\usepackage{xcolor}

\usepackage{CJK}
\usepackage{physics}
\usepackage{array}
\usepackage{multirow}
\usepackage{nopageno}
\usepackage{mathtools}
\usepackage{enumerate}
\usepackage{comment}
\usepackage{soul}
\usepackage[colorlinks=true, linkcolor=red, citecolor=blue]{hyperref}

\usepackage[T1]{fontenc}
\usepackage{verbatim}
\usepackage{xcolor}
\usepackage{pgf,tikz}
\usepackage{mathrsfs}
\usepackage{amsmath,amssymb,textcomp}
\everymath{\displaystyle}

\usepackage{times}
\usepackage{tgheros}

\usepackage[english]{babel}
\usepackage{eurosym}
\usepackage{siunitx}

\usepackage{graphicx}
\graphicspath{{./img/}}
\usepackage{svg}
\usepackage[section]{placeins}

\usepackage{hyperref}
\usepackage[normalem]{ulem}

\newcommand{\esec}{\epsilon_{\text{sec}}}
\newcommand{\ecor}{\epsilon_{\text{cor}}}
\newcommand{\Z}{\text{Z}}

\newcommand{\h}{\text{h}}
\newcommand{\dcr}{CR_d}

\begin{document}

\author{Claudia De Lazzari}
\altaffiliation{These authors equally contributed to this work.\\E-mails: claudia.delazzari@qticompany.com; tecla.gabbrielli@cnr.it}
\affiliation{QTI s.r.l., Largo E. Fermi, 6 - 50125 Firenze, Italy}

\author{Tecla Gabbrielli}
\altaffiliation{These authors equally contributed to this work.\\E-mails: claudia.delazzari@qticompany.com; tecla.gabbrielli@cnr.it}
\affiliation{CNR - Istituto Nazionale di Ottica (CNR-INO), Largo E. Fermi, 6 - 50125 Firenze, Italy}
\affiliation{LENS - European Laboratory for Non-Linear Spectroscopy, Via Nello Carrara, 1 - 50019 Sesto Fiorentino FI, Italy}

\author{Natalia Bruno}
\affiliation{CNR - Istituto Nazionale di Ottica (CNR-INO), Largo E. Fermi, 6 - 50125 Firenze, Italy}
\affiliation{LENS - European Laboratory for Non-Linear Spectroscopy, Via Nello Carrara, 1 - 50019 Sesto Fiorentino FI, Italy}

\author{Francesco Cappelli}
\affiliation{CNR - Istituto Nazionale di Ottica (CNR-INO), Largo E. Fermi, 6 - 50125 Firenze, Italy}
\affiliation{LENS - European Laboratory for Non-Linear Spectroscopy, Via Nello Carrara, 1 - 50019 Sesto Fiorentino FI, Italy}

\author{Domenico Ribezzo}
\affiliation{Università degli Studi di Firenze, Department of Physics and Astronomy, Firenze 50019, Italy
}

\author{Nicola Biagi}
\affiliation{QTI s.r.l., Largo E. Fermi, 6 - 50125 Firenze, Italy}

\author{Nicola Corrias}
\affiliation{QTI s.r.l., Largo E. Fermi, 6 - 50125 Firenze, Italy}

\author{Simone Borri}
\affiliation{CNR - Istituto Nazionale di Ottica (CNR-INO), Largo E. Fermi, 6 - 50125 Firenze, Italy}
\affiliation{LENS - European Laboratory for Non-Linear Spectroscopy, Via Nello Carrara, 1 - 50019 Sesto Fiorentino FI, Italy}

\author{Mario Siciliani de Cumis}
\affiliation{ASI Agenzia Spaziale Italiana - Centro Spaziale ``G. Colombo'', Località Terlecchia, Matera, 75100, Italy}

\author{Paolo De Natale}
\affiliation{CNR - Istituto Nazionale di Ottica (CNR-INO), Largo E. Fermi, 6 - 50125 Firenze, Italy}
\affiliation{LENS - European Laboratory for Non-Linear Spectroscopy, Via Nello Carrara, 1 - 50019 Sesto Fiorentino FI, Italy}

\author{Davide Bacco}
\affiliation{QTI s.r.l., Largo E. Fermi, 6 - 50125 Firenze, Italy}
\affiliation{Università degli Studi di Firenze, Department of Physics and Astronomy, Firenze 50019, Italy
}

\author{Alessandro Zavatta}
\affiliation{QTI s.r.l., Largo E. Fermi, 6 - 50125 Firenze, Italy}
\affiliation{CNR - Istituto Nazionale di Ottica (CNR-INO), Largo E. Fermi, 6 - 50125 Firenze, Italy}

\title{Free-space time-bin encoded quantum key distribution from near- to mid-infrared wavelengths}

\begin{abstract}
Quantum technologies play a central role in establishing new ways of quantum-secured communication. We investigate Free-Space Quantum Communication and explore the advantage of implementing Quantum Key Distribution (QKD) with weak coherent states produced by a light source in the Mid-Infrared (>\SI{3}{\micro \meter}). We simulate time-bin encoded quantum key distribution and demonstrate that a free-space QKD link operating in the Mid-infrared outperforms configurations based on conventional near-infrared wavelengths under various weather scenarios, with a particular significance in adverse meteorological conditions.
\end{abstract}

\maketitle

\section*{Introduction}
\label{sec:Introduction}
\begin{figure*}[t]
    \centering
    \subfigure[]{\includegraphics[width=0.29\textwidth]{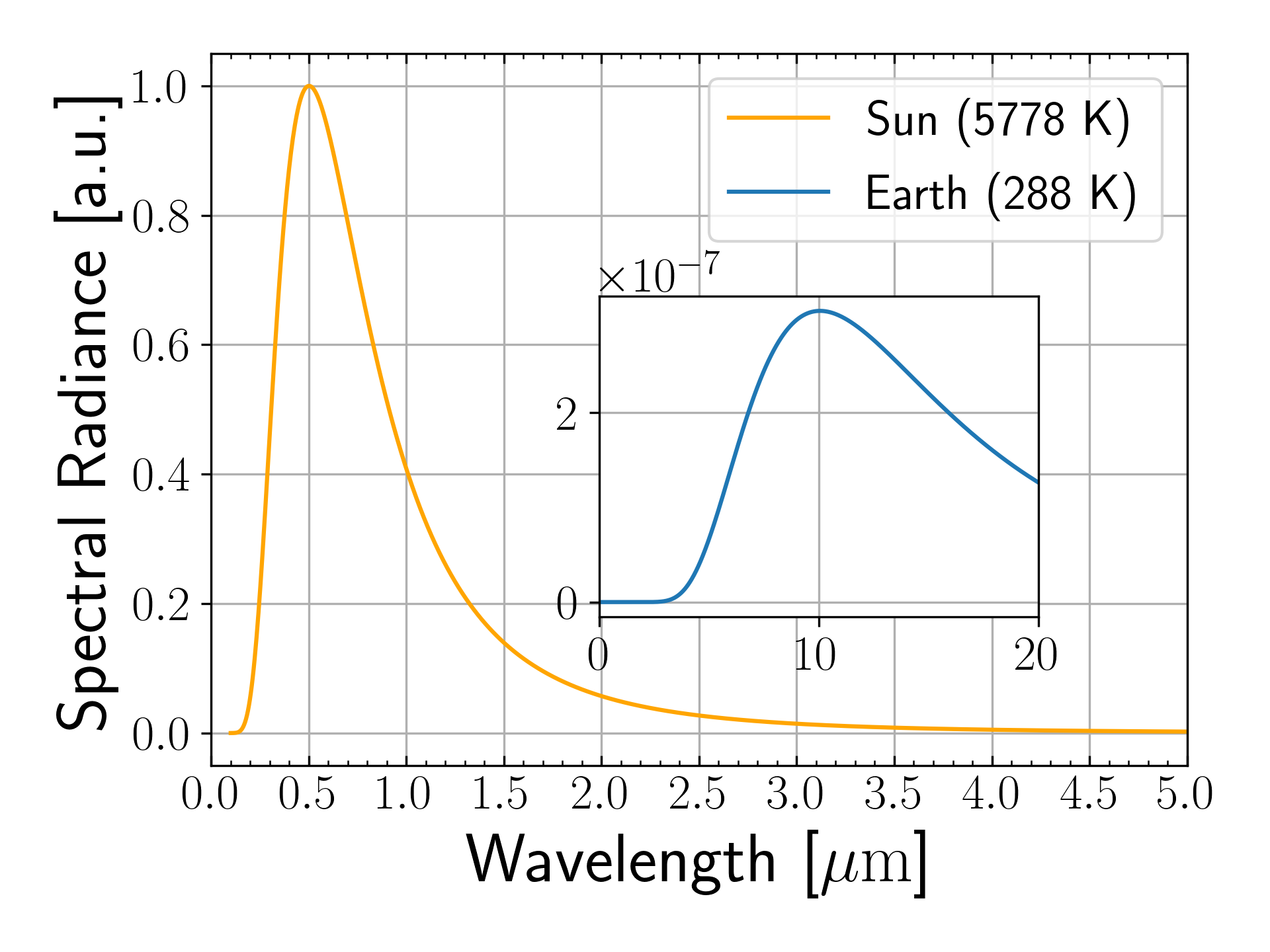}}
    \subfigure[]{\includegraphics[width=0.7\textwidth]{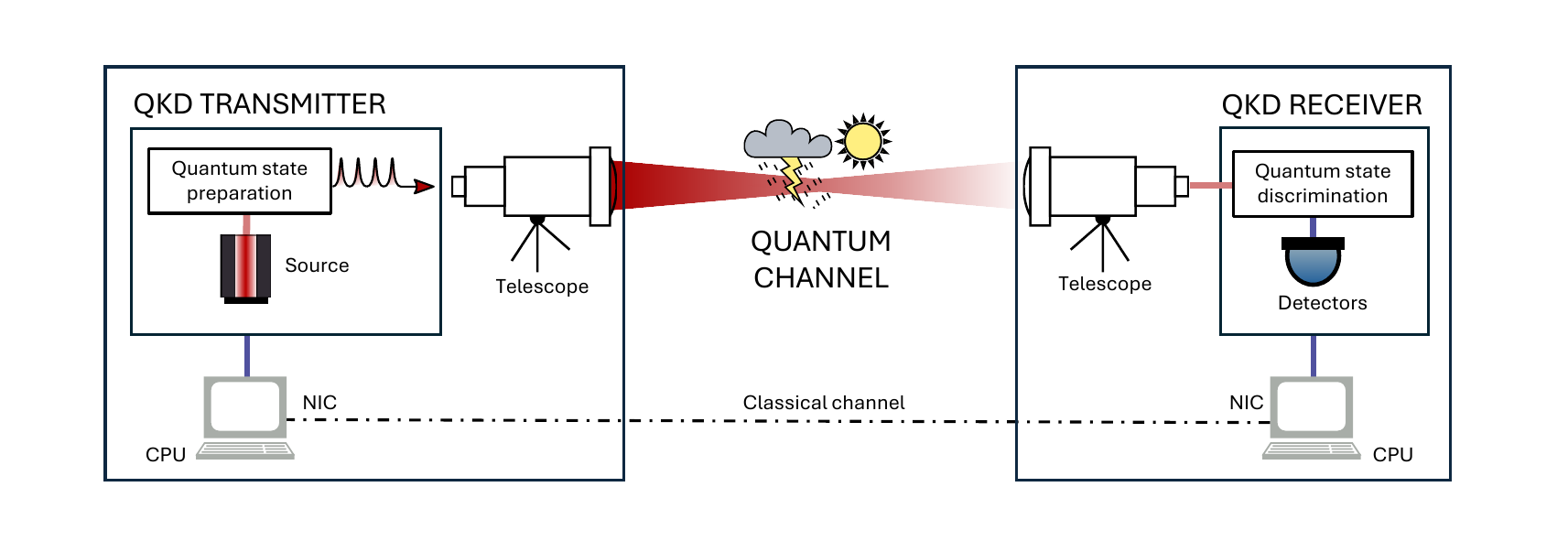}}
    \caption{{(a) Black-body Spectral Radiation in function of the wavelength, normalized to the peak maximum of the the Sun. The background noise due to the black-body
radiation of the Sun (temperature $5778 $ K) is peaked at $\lambda \sim 500$ nm and suppressed above \SI{3}{\micro m}. For Earth (temperature $288$ K), it is peaked at $\lambda \sim 10$ \SI{}{\micro m} and suppressed below \SI{5}{\micro m}. (b) Graphical schematics of the free-space quantum communication link. From left to right: the transmitter composed of a laser source, the quantum state preparation stage, a telescope and a Central Processing Unit (CPU) (using a Network Interface Card (NIC) to set the classical channel); the free-space quantum channel affected by different weather conditions; the receiver composed of a telescope, a quantum states discrimination stage, the detectors, and a CPU (using a NIC to set the classical channel). }}
    \label{fig:schematics}
\end{figure*}
Data transmission and storage safety are pivotal in the worldwide effort to build complex communication infrastructures \cite{simon2017towards,Diamanti2017,conti2024satellite}. In this framework, quantum technologies play a central role in establishing new ways of communications whose security is guaranteed by the principles of quantum mechanics~\cite{bennett2014quantum, Ekert1992, gisin:2007, Scarani2009, pirandola2020advances}. Many experiments demonstrating fiber-based and free-space Quantum Key Distribution (QKD) have been carried out, providing a strong technological advancement in this field regarding both components and communication protocols. Although there is a vast literature studying these topics in the near-infrared (near-IR, i.e. 0.78-3 \SI{}{\micro \meter}~\cite{footnote5}) bandwidth~\cite{Buttler2000,hughes2002practical,Rarity2002,Bonato2009,Liao2017,Ren2017,Ji:17,Liao2017,liao2017long, yin2020entanglement, cao2020long, berra2023modular,cai2024free}, there are few studies that explore the possibility of implementing quantum secure communications in the mid-infrared (mid-IR, i.e. 3-50 \SI{}{\micro \meter}) of the electromagnetic spectrum. 
In particular, previous studies regarding free-space links~\cite{Manor:03,Temporao2008,Grabner:14, Su:18,Corrigan:09} show that in case of adverse meteorological conditions, mid-IR can outperform the most commonly used optical wavelengths, as analyzed in this paper.
In some high-transparency spectral windows, atmospheric absorption for mid-IR radiation (e.g. $\approx$ \SI{1.9e-3}{\deci\bel/\kilo\meter} around \SI{4}{\micro m})  becomes comparable to telecom wavelength ($\approx$  \SI{1e-3}{\deci\bel/\kilo\meter} around \SI{1.6}{\micro m}) ~\cite{footnote2,Ycas2018,dello2022advances,elefante2023recent,limid:2025}. \\
Among free-space communication channels, mid-IR experiences reduced  
Rayleigh scattering from atmospheric molecules~\cite{footnote1}, greater resilience to turbulence and severe weather conditions~\cite{Manor:03,Su:18,Hao:17}, and lower background noise, which makes it suitable for daytime employment.
The background noise \cite{Loudon:2000} due to
Sun and Earth black-body radiations is suppressed above \SI{3}{\micro m}, and below \SI{5}{\micro m}, respectively; see Fig. \ref{fig:schematics}(a).
As a consequence, several works on the feasibility and performance evaluation of practical free-space applications of mid-IR classical communication systems have been released over the last years \cite{Pang:2020,pang:22,Marsland:24, Didier:23}. 
In parallel, triggered by possible quantum communication and/or quantum sensing applications, a few studies aimed at designing and developing specific mid-IR technologies have recently been published~\cite{Gabbrielli:2021,trombettoni2021quantum,Gabbrielli:2022,Marschick:2024,Gabbrielli:2025}.\\ 
Despite the growing interest in quantum applications within the mid-IR, this spectral region still experiences a technological gap compared with the near-IR one in both non-classical state generation \cite{Gabbrielli:2022} and detection performance (photovoltaic detectors reported in the literature typically exhibit quantum efficiencies of around 40\% \cite{Gabbrielli:21}).
Moreover, adequate and high-performance single-photon detectors are currently underdeveloped even though some studies are already available in the literature \cite{marsili2012efficient,pan2022mid,chang2022efficient,CHEN2021965}.
Therefore, the development of QKD implementations remains a relatively unexplored path, as currently only a few works analyze the feasibility of QKD with mid-IR sources and polarization encoding of the quantum states \cite{Temporao2008, limid:2025}. Preliminary characterization is available in the literature \cite{temporao2006mid}. 
In this scenario, focused simulations of the QKD protocol assuming a mid-IR setup compared to the near-IR standard are relevant to set key points and target performances that can help to facilitate and accelerate the required technological optimization.\\
{Polarization encoding is generally the most common choice in implemented Free-Space Quantum Communication (FSQC) links \cite{hughes2002practical, Liao2017,liao2017long, yin2020entanglement, cao2020long, berra2023modular,cai2024free, limid:2025}.
However,  recent studies have 
demonstrated  time-bin encoding as a convenient way for implementing horizontal QKD links with near-IR light \cite{jin2019genuine, merolla2023high, jennewein2023time, Cocchi:25}.} \\
{Here, we simulate an efficient time-bin QKD link, comparing the use of near-IR (telecom and 800~nm) and mid-IR sources over free-space channels in clear, rainy, and foggy weather. 
We show that the mid-IR source offers substantial advantages in terms of channel loss, including greater resilience to adverse weather conditions, making it a strong candidate for advanced free-space QKD implementations.
We also analyze QKD receiver setups with different single-photon detectors.}

\section{Methods and Technologies}
\label{sec:Methods}
In this work, we adopt a QKD protocol following a common formalism independent of the wavelength. To compare the two different regimes of mid-IR and near-IR radiation, we first need to introduce the QKD protocol and then the main technical parts of an FSQC channel. 
In Fig. \ref{fig:schematics}(b) we show a schematic of the fundamental components of a FS QKD link, divided for simplicity into QKD transmitter, noisy and lossy channel, and QKD receiver. 

\subsection{Quantum Key Distribution}
\label{subsec:QKD}
\begin{figure}
    \centering
    \includegraphics[width=0.5\textwidth]{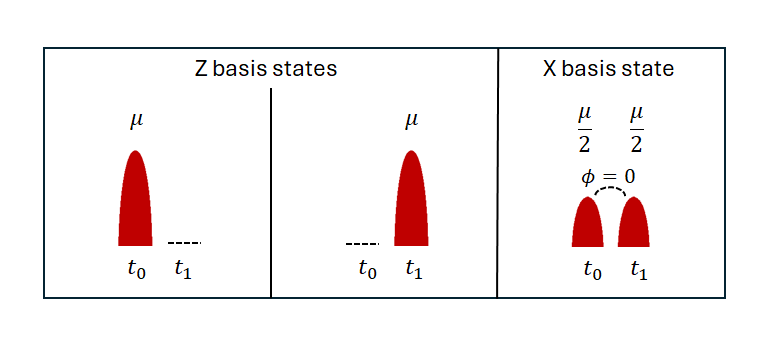}
    \caption{Quantum states used in the 3-states BB84 protocol with time-bin encoding. States of the Z basis have signal in the first ($t_0$) or the second ($t_1$) time-bin, respectively. The state in the X basis is a superposition state with zero relative phase. The signal intensity is $\mu$ and $\mu/2$ for the states in the Z and X basis, respectively.}
    \label{fig:tre_stati}
\end{figure}

QKD allows the distribution of a symmetric key between distant parties, based on the laws of quantum physics. Each QKD protocol
requires the preparation, transmission, and measurement of certain quantum states, and finally, the processing of the measurement outcomes. 
Indeed, the protocol usually consists of two phases. In the first one, quantum states are 
exchanged through a completely public quantum channel.
In a second part, the outputs of the quantum measurements are analyzed and elaborated to distill a secure classical key. To achieve this, the parties exchange classical information through an authenticated classical channel. The classical channel can be fiber-based or can use Network Interfaces as shown in Fig. \ref{fig:schematics}(b). 
The produced key is unconditionally secure, since the protocol is resilient against any type of admissible attack (quantum and classical), and can be used as a symmetric secure key in cryptographic applications. \\
The adopted protocol is the discrete-variable efficient 3-state BB84, in the finite-key regime, with 1-decoy state method \cite{Ma:2005, lo2005efficient,tomamichel2012tight, lim2014concise, rusca2018finite}, and time-bin encoding. 
The transmitter, Alice, prepares quantum states as phase-randomized laser pulses, attenuated to the single-photon level. The states have different intensities as prescribed by the decoy-state method. 
The possible states are three, in two mutually unbiased bases, Z and X. 
Given two temporal intervals, $t_0$ and $t_1$, the states of the Z basis are defined by the time of occupation of the signal, first and second, respectively. The state prepared in the X basis 
is a superposition of the Z basis states, with zero relative phase ($\phi = 0$) between the two time bins, see Fig. \ref{fig:tre_stati}. 
Alice randomly chooses to send states prepared in the Z or X basis.
At the receiver side (Bob), the incoming signals are measured by single-photon detectors (SPD). 
For the Z basis, an SPD measures the time of arrival of the photons. 
The states in the X basis are measured using an unbalanced interferometer that provides the phase measurements.
After the quantum communication, a post-processing stage is applied to the measured results, and a final secure key is distilled. In the adopted protocol, only the Z basis states encode the secure bits, while the X basis detections are used for the security analysis. From a block of $n_\Z$ sifted bits,
a secure key of length $skl< n_\Z$ can be produced. In this work, we use the secure key rate ($skr$) as a quality factor to compare the different QKD implementations. The rate $skr$ is obtained by the ratio between $skl$ and the time necessary for exchanging $n_\Z$ sifted bits. In detail, for the 1-decoy 3-state BB84 protocol in the finite-key regime, $skl$ is bounded by \cite{lim2014concise,rusca2018finite}:
\begin{equation}
skl\leq s_{\Z,0}^\ell+s_{\Z,1}^\ell\left[1-\h(\phi_{\Z}^u)\right]-\left[f(q_\Z,\ecor)+g(\esec)\right]
    \label{eq:sec length}
\end{equation}
with $s_{\Z,0}^\ell$ and $s_{\Z,1}^\ell$ being the lower bounds for the vacuum and the single-photon events, $q_\Z$ the Z basis quantum bit error rate (QBER), and $\phi_{\Z}^u$ the upper bound of the Z phase-flip error rate; $\phi_{\Z}^u$ is inferred from the X basis QBER, due to the properties of mutual unbiased bases. The function $\h(x)=-x\log_2(x)-(1-x)\log_2(1-x)$ is the binary entropy, and the quantity $\h(\phi_{\Z}^u)$ represents the maximum information that an adversary can extract introducing $\phi_{\Z}^u$ phase-flip errors. Removing this quantity corresponds to the privacy amplification process, costing (including finite-size key effects) $g(\esec)=6\log_2\left(19/{\esec}\right)$, with the secrecy parameter $\esec=10^{-15}$. 
The amount of information disclosed in the error correction and verification stage is $f(q_\Z,\ecor)=1.12 \ n_\Z \ \h(q_\Z)+\log_2\left(2/{\ecor}\right)$ with $\ecor=10^{-15}$ correctness parameter. The length of the raw key is set to $n_\Z = \SI{e8}{bit}$.\\

\noindent 
In the following, we give a description and an overview of the current technological level of the main components (i.e., transmitter, channel model, and receiver), corresponding to each element of Fig. \ref{fig:schematics}(b), where we consider an Earth point-to-point QKD link. 

\subsection{Transmitter} 
\label{subsec:Transmitter}
The key parts on the transmitter side are the laser source, the modulators and attenuators, together with the necessary electronics, and the telescope.\\

\paragraph{Laser sources.}
Commercial high-performing laser sources are available both in the near-IR and in the mid-IR. \\ In the near-IR, diode lasers are typically used in a standard QKD setup.
Concerning the mid-IR, a standardized QKD system cannot be considered established yet. However, the mid-IR can benefit from high-performance coherent sources, already tested in classical communication. Among them, Quantum Cascade Lasers (QCLs)~\cite{Faist:1994} 
emerges as high-performance chip-scale semiconductor devices allowing for high-coherence radiation and fast modulation~\cite{Hinkov:2019,Hillbrand:2019, Pirotta:2021}. 
Recently, QCLs have been exploited in the 3-\SI{5}{\micro m} mid-IR transparency window for proof of principle free-space classical communication link in a controlled laboratory environment, reaching a data rate of a few Gbps when used in direct modulation scheme~\cite{Pang:2020,pang:22,Marsland:24,mikolajczyk2014overview}.\\
Alongside QCLs, Interband Cascade Lasers (ICLs) have been gaining attention in the last year as an alternative laser source directly emitting in the mid-IR~\cite{Yang:1995}. These devices have been exploited in a controlled laboratory environment for proof-of-principle free-space classical communication links. Data rates above \SI{10}{\giga\bit/\s} (\SI{12}{\giga\bit/\s} with an on–off keying scheme and \SI{14}{\giga\bit/\s} with a 4-level pulse amplitude modulation scheme have been achieved \cite{Didier:23}). ICLs are preferable in terms of energy consumption, however they have a lower output power \cite{Gabbrielli:2025} and in direct modulation schemes they have a limited bandwidth (tens of \SI{}{\giga\hertz}) compared to QCLs (hundreds of \SI{}{\giga\hertz}) due to their slower intrinsic dynamics \cite{Grillot:2025}. Regarding intensity noise features, ICLs benefit from a reduction of excess noise compared to QCLs \cite{Gabbrielli:2025}.  
Moreover, sources like QCLs and ICLs are potential candidates for integrated non-classical light emitters, thanks to the high third-order nonlinearity of their active region~\cite{levenson:1985}, responsible also for their peculiar frequency comb emission, based on four-wave-mixing process \cite{Faist:2016,Riedi:2015,Friedli:2013,Meyer:2020}. 
The direct comb emission that characterizes these types of mid-IR sources~\cite{Faist:2016, Meyer:2020} enables the exploitation of multiplexing in communication protocols. Moreover, their intermodal beat note signal, which falls in the 5-20 \SI{}{\giga\hertz} range, can be easily interfaced with microwave systems for data encoding and transmission~\cite{Piccardo:2019a,Corrias:2022}.
Analogously to the near-IR case, mid-IR laser sources fully satisfy the demands in terms of stability and spectrum broadness (e.g. few hundred Hz or below \cite{Bartalini:2010,cappelli:2012} in the case of QCLs, and around tens of kHz for ICLs \cite{li202230}) of a QKD transmitter. \\

\paragraph{Modulators.}
In contrast, while in the near-IR efficient modulators with typical bandwidths of tens of \SI{}{\giga\hertz} are commercially available, high-speed mid-IR modulators are still an ongoing research topic, primarily \cite{herrmann2020modulators,XuDong2023}. 
As reported in refs.\cite{Jafari:2023, Grillot:2025}, different types of modulators have been demonstrated in the mid-IR region. As an example, Germanium-on-Silicon (Ge-on-Si) modulator with an on off keying modulation up to \SI{60}{\mega\hertz} has been proven at \SI{3.8}{\micro m}, \cite{Li:19}. Moreover, modulator based on the same materials have been proved capable of reaching modulation of \SI{1}{\giga\hertz} \cite{nguyen20221} and above (\SI{60}{\giga\hertz} \cite{barzaghi2022modelling}) at longer wavelength (above \SI{5}{\micro m}). Moreover, a recent study numerically investigated a novel structure based on Graphene covered on SiO$_2$ gratings, where a modulation speed up to \SI{1.5}{\tera\hertz} with a modulation depth of 100\% is predicted in the 1-90 \SI{}{\tera\hertz} spectral region \cite{Marsland:24}. Another modulation approach pursued with QCLs consists in RF-injection modulation scheme, where modulation at tens of \SI{}{\giga\hertz} (both electrical and optical) has been reported, albeit with limited modulation depth \cite{Hinkov:16}.
\\ Despite some promising results, to mitigate the technological gap in modulation speed affecting the mid-IR, we provide in this work the secure key rate per pulse, which is independent of the state preparation rate and therefore of the modulation capability. \\

\paragraph{Telescope.} Finally, despite its longer wavelength compared to the near-IR band, mid-IR light can be confined in Gaussian beams of a specific size, which can be shaped and directed with very high precision over kilometer-scale distances using optimally designed systems \cite{Abadi2019}. As in the near-IR region, FSQCs can be implemented using mid-IR sources by exploiting standard collimating telescopes \cite{Manor:03,Kurtsiefer2002} as well as Cassegrain or Gregorian telescopes \cite{Hemmati:2003}. For a typical urban communication scenario, where the transmission distances are on the order of tens of kilometers, geometrical losses can be minimized by using two identical telescopes for the transmitter and the receiver.

\subsection{Receiver}
\label{subsec:Receiver}
On the receiver side, in the proposed simulation, we assume that the telescope has the same dimensions as that on the transmitter side. 
Another key component considered in the simulation is the {single-photon} detector, since its parameters -- mainly efficiency and dark count rate -- affect the quality of quantum states measurements.
The near-IR region benefits from high-speed low-noise single-photon detectors optimized for quantum measurements, such as Indium Gallium Arsenide (InGaAs) avalanche photodiodes and superconducting nanowire single-photon detectors (SNSPDs). The latter, in particular, offer high detection efficiency (over 98\%), along with minimal timing jitter and hold-off times as short as a few tens of picoseconds \cite{natarajan2012superconducting,reddy2020superconducting,chang2021detecting,salvoni2022activation}. By integrating a filtering stage directly on the nanowire, it is also possible to achieve dark count rates below \SI{1}{\hertz} \cite{natarajan2012superconducting,salvoni2022activation}.
In contrast, mid-IR technology is still relatively immature, referring to single-photon detection \cite{marsili2012efficient,pan2022mid,chang2022efficient,CHEN2021965, dello2022advances}.  
In this work, we focus on discrete variables, which require single-photon detection modules. Recently, research on {superconducting} single-photon detectors has also extended to the mid-IR region. Preliminary studies have been conduct reaching detection efficiency of $\eta_{\text{det}}=0.05$ and a dark count rate of {$CR_d=100$}~\SI{}{\hertz} \cite{CHEN2021965}.
As an alternative, up-conversion detection scheme from mid-IR to near-IR can be evaluated for the benefit of higher detection efficiency, such as done e.g. in Ref. \cite{temporao2006mid}. In the following simulation, we analyze both of these detection alternatives. 

\subsection{Channel Model and Losses} 
\label{subsec:Channel}
During propagation, quantum signals are affected by  distance-dependent losses that arise from various factors such as Gaussian beam divergence, atmospheric absorption, and adverse weather conditions.
\\
\paragraph{Geometrical losses.}
For both mid-IR and near-IR signals, path losses are primarily due to the divergence of Gaussian beams, and can be modeled as:
\begin{equation} 
    L_{\mathrm{geo}} = 10\log_{10} \left(\frac{S_{d}}{S_{\mathrm{Rx}}} \right) \text{,}
\end{equation}
where $S_d$ is the surface area of the transmitted beam at the communication distance $d$, calculated according to the Gaussian beam propagation (see Supplementary material \ref{appx:Suppl}), 
and $S_{\mathrm{Rx}}$ is the receiver capture surface. Denote by $w_{0}$ the beam waist radius, defined as the radius at which the beam intensity falls to $1/e^2$ of its maximum value. 
Assuming that both telescopes (transmitter and receiver) have equal radius  $r_{\mathrm{Rx}}=\sqrt{2} w_{0}$ and that the beam waist $w_{0}$ is placed at the Rayleigh distance $Z$ from the transmitter (i.e. half way between transmitter and receiver, as in ref. \cite{Corrias:2022}), we can get the following equation describing the ratio  
between the area of the beam 
and the area of the receiver/transmitter telescope: 
\begin{equation}
    L_d=\frac{S_d}{S_{\mathrm{Rx}}} =\frac{1}{2}\left[1+\left(\frac{d - Z}{Z}\right)^2\right], \quad Z =  \frac{\pi \ r_{\mathrm{Rx}}^2}{2 \lambda}
\end{equation}
Therefore, the geometrical loss is given by
\begin{equation}
    L_{\mathrm{geo}} =
    \begin{cases}
        10 \log_{10}(L_d) & d > 2 Z, \\
        0 & \text{otherwise}.
    \end{cases}
\end{equation}
Notice that it is set to zero whenever $S_d \leq S_{\mathrm{Rx}}$.\\
Besides the geometrical attenuation -- that naturally arises from the divergence of a Gaussian beam with distance -- the design of an optical transmitter or receiver must also account for factors such as pointing stability, tolerance to atmospheric turbulence, and aberration correction. 
We are not considering these aspects in our analysis, but we remark that they need to be taken into account in case of practical application, as they impact both the overall loss budget and the compactness of the system in a real implementation.\\
\paragraph{Scattering and absorption.}
The absorption and scattering losses can be modeled via the transmittance $\tau_{\mathrm{eff}}$ given by:
\begin{equation}
    \tau_{\mathrm{eff}} = 10^{-\gamma_d/ 10} ,
\end{equation}
where $\gamma_d = \gamma \times d$, with $\gamma$ is the specific attenuation in \SI{}{\deci\bel/\kilo\meter} of the considered loss mechanism and $d$ is the propagation distance in \SI{}{\kilo\meter}. In this work, these losses are attributed to atmospheric absorption and weather conditions (in particular rain and fog). 
We used the HITRAN database to choose the local minima of the atmospheric molecular absorption at the used wavelengths \cite{footnote2}.
In the next section, we introduce the additional losses due to specific weather conditions, air molecules' dimensions, and turbulence. \\

\paragraph{Non-optimal weather conditions.}
\label{subsec:weather_conditions}
For FSQC, it is critical to consider weather conditions as they act as additional losses,  significantly affecting the communication \cite{Kaushal2018}. \\
Denote the mid-IR and near-IR attenuation by $\gamma_{\mathrm{MIR,r}}$ and $\gamma_{\mathrm{NIR,r}}$ (\SI{}{\deci\bel/\kilo\meter}), where $r$ is the radius of the particles encountered by the quantum states along the free-space path. \\
Weather conditions are characterized in terms of \emph{Mie scattering}, i.e. the case $r \approx \lambda$, whose losses are empirically given by:
\begin{equation}
\label{losses_rain_mw}
\gamma_{\mathrm{\lambda,r}} =  10\,\log_{10}\left(e\right)\,\frac{3.91}{V}\, \left(\frac{\lambda}{550} \right)^{-p},
\end{equation}
where $\lambda$ (\SI{}{nm}) corresponds to either mid-IR or near-IR wavelength, $V$ (\SI{}{\kilo\meter}) is the visibility \cite{footnote6}, and $p$ is a scattering coefficient, depending on the visibility: 
\begin{equation}
    p = \begin{cases}
    1.6 & V > \SI{50}{\kilo\meter}, \\
    1.3 &  \SI{6}{\kilo\meter}\leq V \leq \SI{50}{\kilo\meter}, \\
    0.585 \ V^{1/3} &  V < \SI{6}{\kilo\meter}.
    \end{cases}
\end{equation}
Notice that the visibility entirely determines the attenuation when considering the Mie scattering.
A clear weather scenario is usually associated with a visibility higher than \SI{20}{\kilo\meter}, whereas fog is associated with a visibility below \SI{1}{\kilo\meter}.\\
Rain causes additional attenuation of the beam, which is wavelength independent. The attenuations $\gamma_{\mathrm{MIR,r}}$ and $\gamma_{\mathrm{NIR,r}}$ (\SI{}{\deci\bel/\kilo\meter}) are modeled as: 
\begin{equation}
\label{losses_rain_mid-IR}
\gamma_{\mathrm{MIR,r}} = \gamma_{\mathrm{NIR,r}} = 1.076\, R^{0.67}, 
\end{equation}
where $R$ (\SI{}{\milli\meter/\hour}) is the rain rate. \\
In the simulations, we compare mid-IR and near-IR sources in clear, foggy, and rainy weather, which impact both frequency domains. \\
In the case $r < \lambda$, the scattering process is classified as \emph{Rayleigh scattering} and must be taken into account. In this regime, the attenuation coefficient is given by: 
\begin{equation}
    \gamma_{\mathrm{\nu,r}} = \frac{\nu^4}{9.26799 \times 10^{18}-1.07123 \times 10^{9} \ \nu^2},
\end{equation}
with $\nu=10^4/\lambda$ being the wavenumber ($\text{cm}^{-1}$).
\\
Finally, the effect of turbulence is considered, including the contribution of scintillation:
\begin{equation}
    \gamma_{\mathrm{\lambda,r}} = 2 \ \sqrt{
1.23 \ (k ^{7 / 6}) \ c_{\text{n}}^2 \ {(d^{11 / 6})},
}
\end{equation}
where $k = 2 \pi / \lambda$ is the wavenumber and $c_{\text{n}}^2 = 10^{-14}$ is the refractive index structure parameter.
\section{Results}
\label{sec:Results}
\begin{figure*}[t]
    \centering
    \subfigure[]{\includegraphics[width=0.40\textwidth]{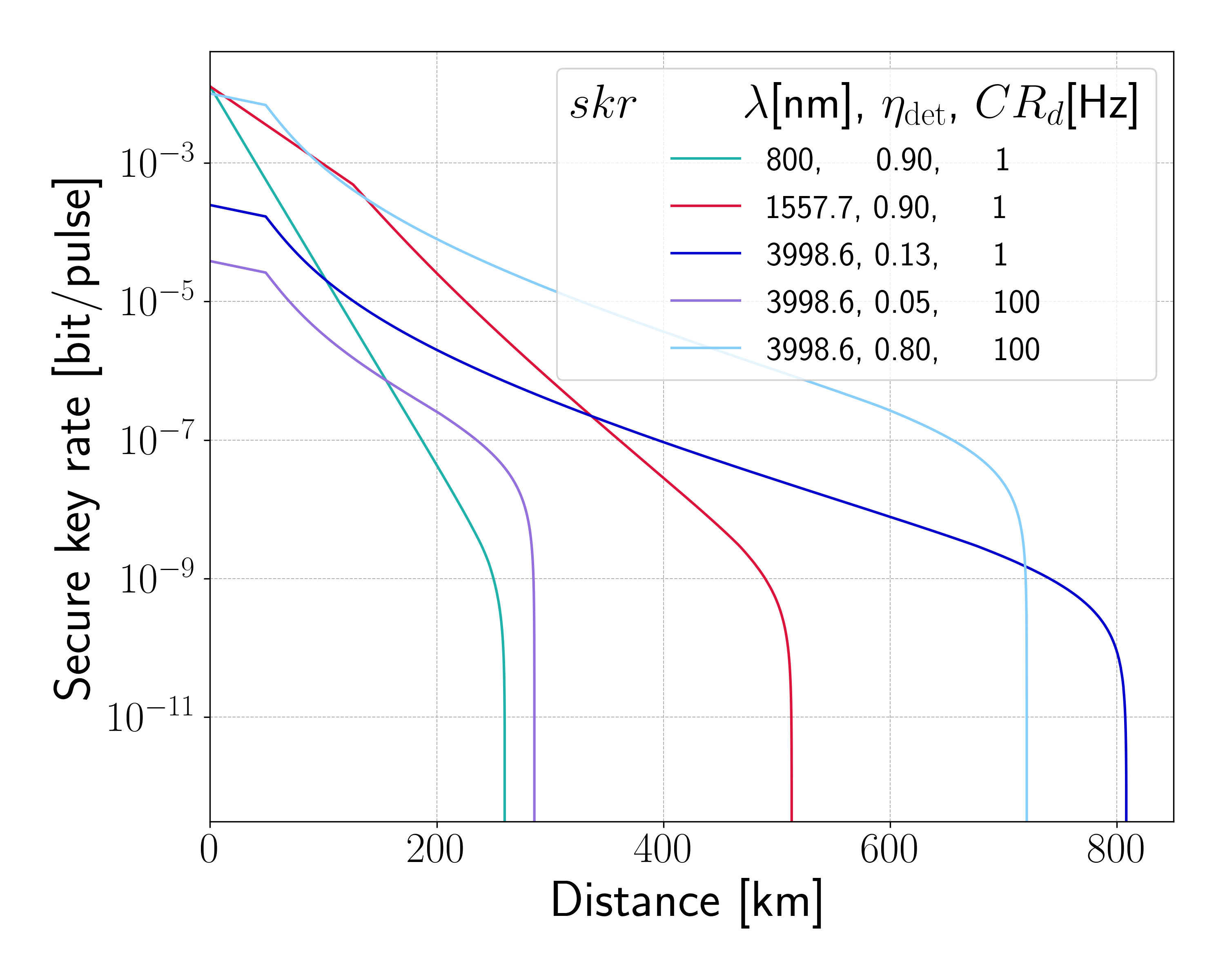}}
    \subfigure[]{\includegraphics[width=0.40\textwidth]{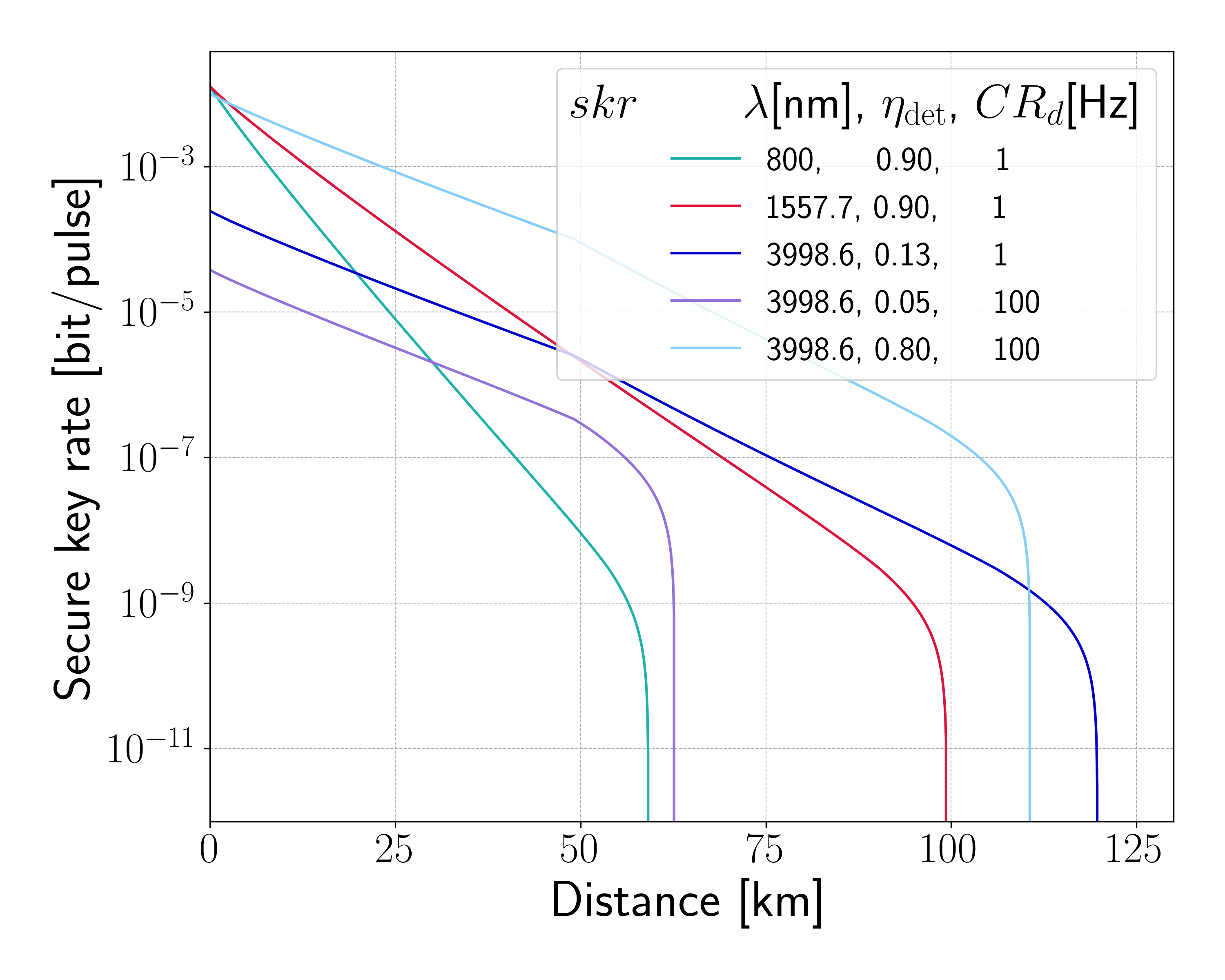}}
    
    \subfigure[]{\includegraphics[width=0.40\textwidth]{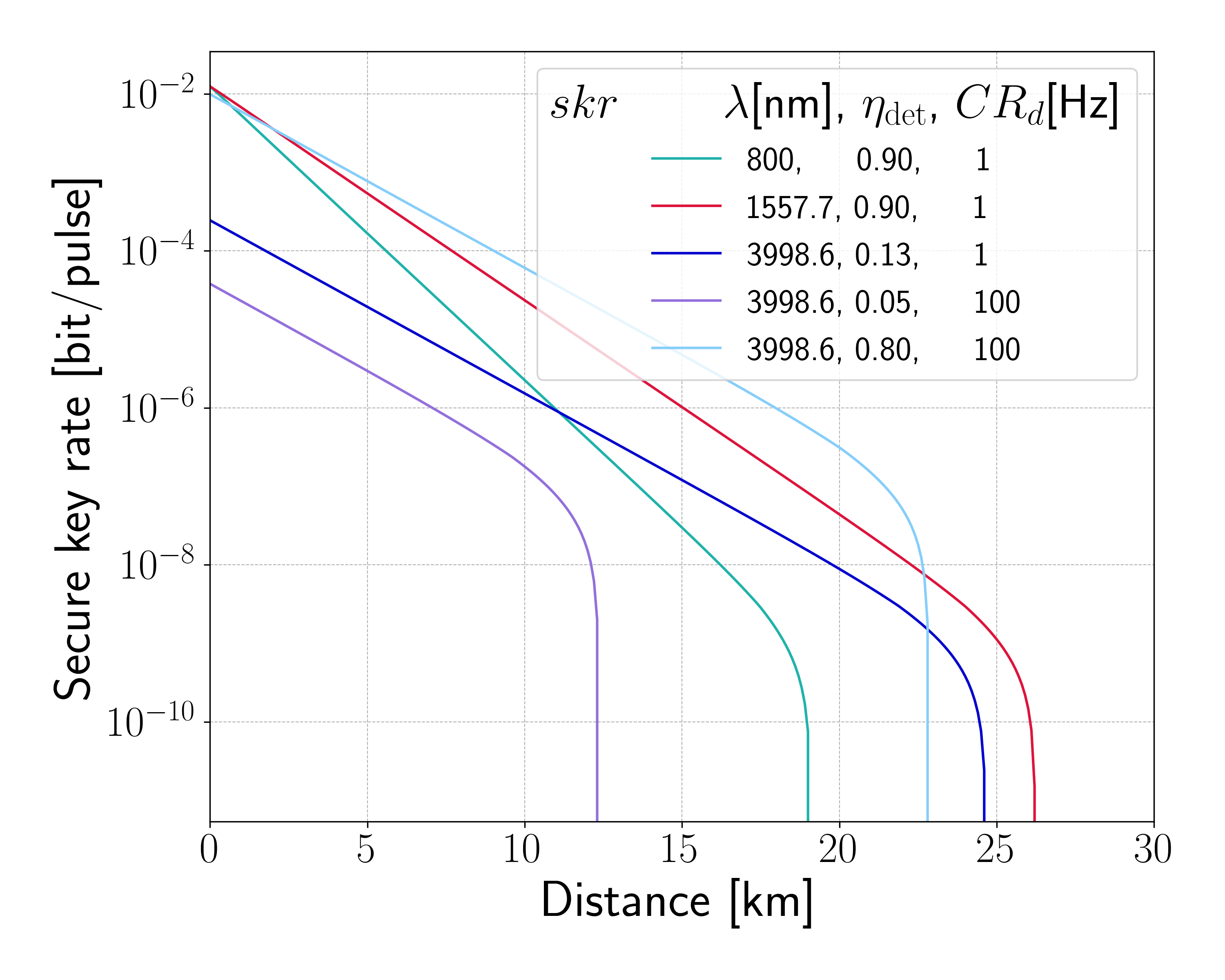}}
    \subfigure[]{\includegraphics[width=0.40\textwidth]{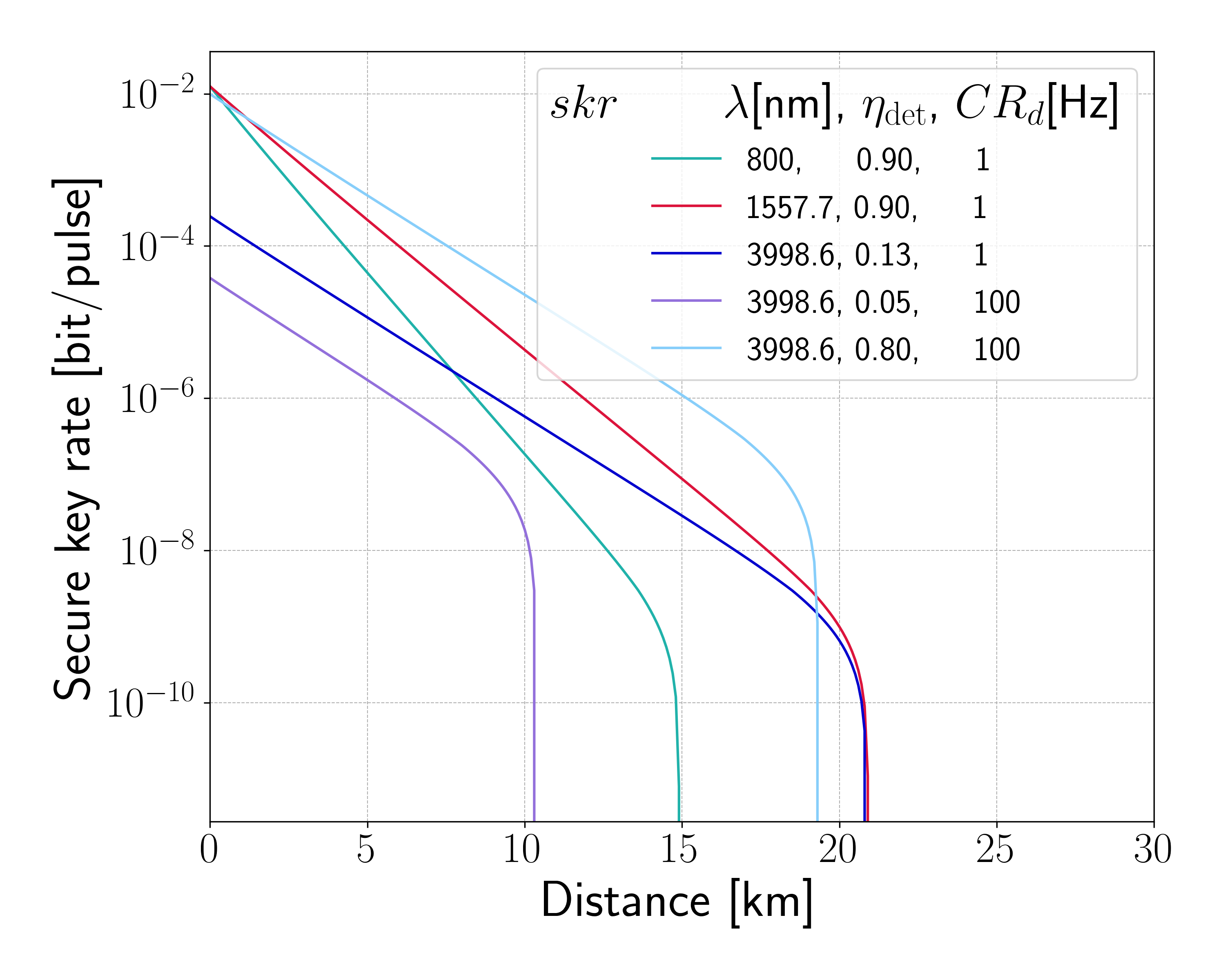}}
    
    \subfigure[]{\includegraphics[width=0.40\textwidth]{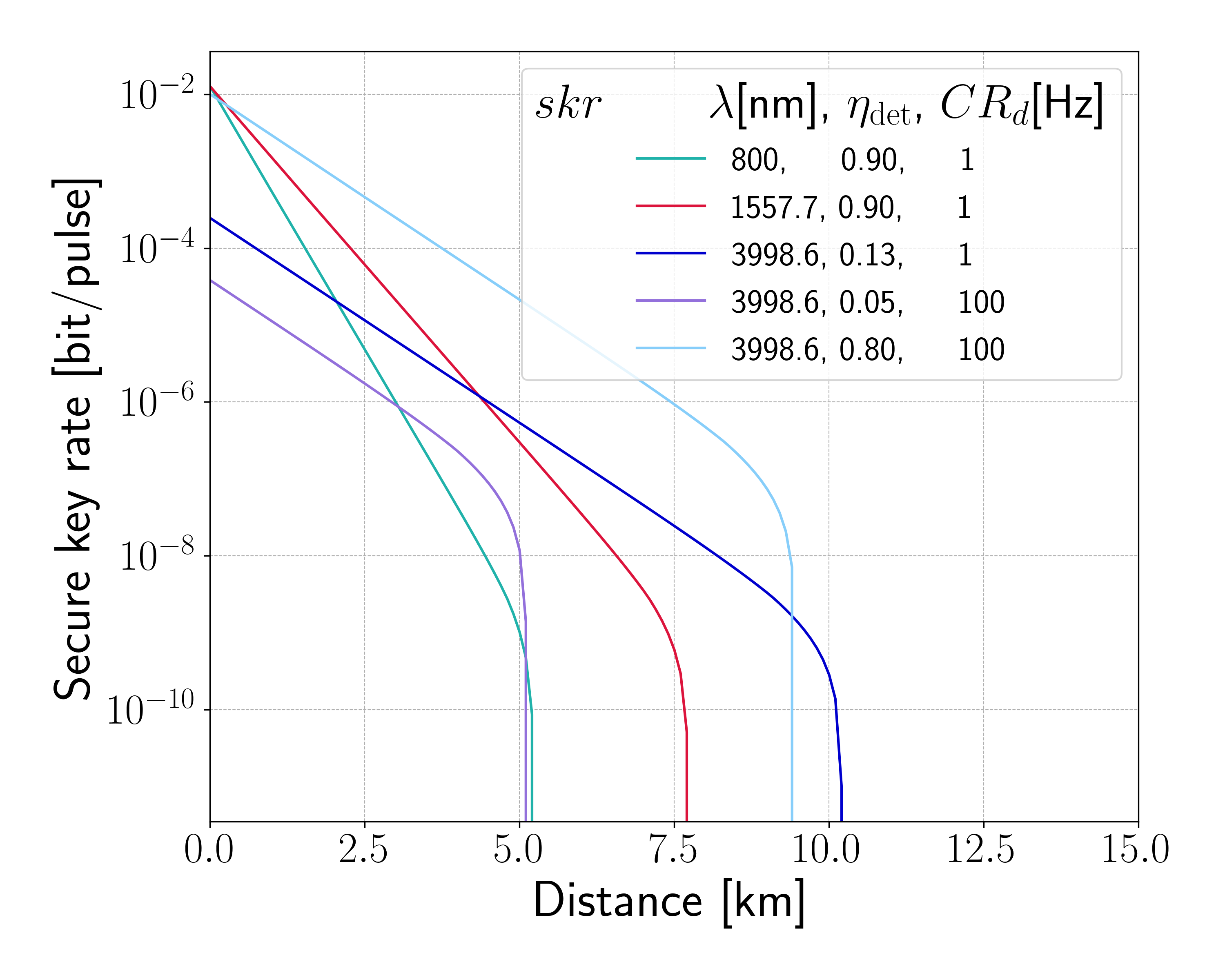}}
    \subfigure[]{\includegraphics[width=0.40\textwidth]{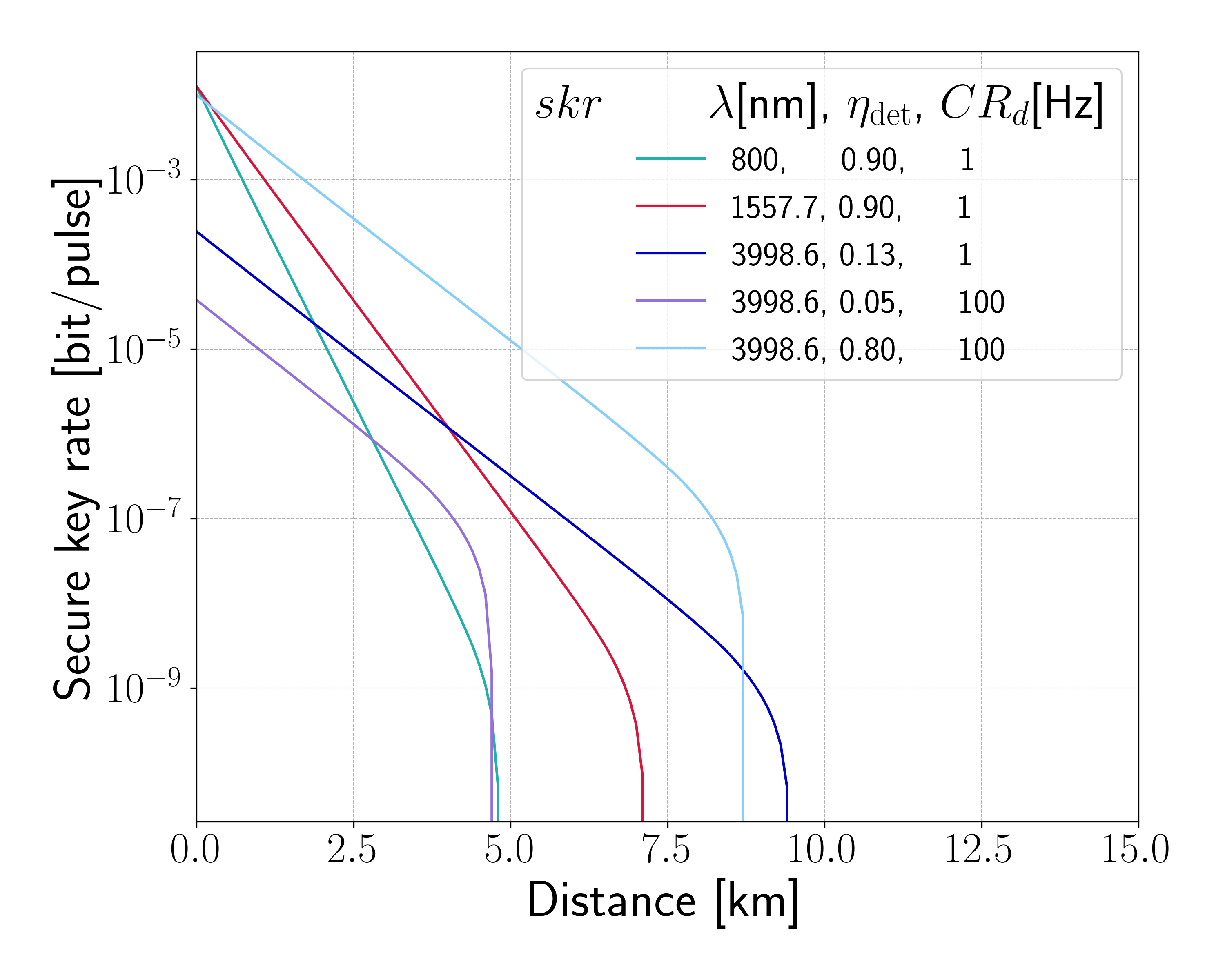}}
    \caption{Comparison of secure key rate per pulse in function of the link length in the weather conditions: (a) clear weather (V = \SI{40}{\kilo\meter} visibility); (b) clear weather with turbulence; (c) rain (R =  \SI{2.5}{\milli\meter/\hour} of rainfall rate associated with a visibility of \SI{6}{\kilo\meter}); (d) rain with turbulence; (e) fog (V = \SI{1}{\kilo\meter} visibility); (f) fog with turbulence.
    Two traces represent the $skr$ generated by near-IR setups, simulated with the same detectors (efficiency $\eta_{\text{det}}=0.9$, dark count rate $\dcr=1$\SI{}{\hertz}): (green) $\lambda =800$ \SI{}{\nano\meter}; (red) $\lambda =1557.7$ \SI{}{\nano\meter}. 
    Three traces represent the $skr$ generated by mid-IR source $\lambda =3998.6$ \SI{}{\nano\meter} and different measurement setups: (blue) with optimized up-conversion to near-IR (efficiency $\eta_{\text{det}}=0.9$, dark count rate $\dcr=1$\SI{}{\hertz}); (violet) realistic detector operating in mid-IR (efficiency $\eta_{\text{det}}=0.05$, dark count rate $\dcr=100$\SI{}{\hertz}); (cyan) optimized detector operating in mid-IR (efficiency $\eta_{\text{det}}=0.80$, dark count rate $\dcr=100$\SI{}{\hertz}).
    }
    \label{fig:skr}
\end{figure*}
The FSQC simulation model is thus constructed by combining the channel model -- which depends both on the wavelength and the selected weather condition -- with the QKD protocol and setup, see Sec. \ref{sec:Methods}. 
For each considered wavelength, we simulate the secure key rate per pulse under several weather conditions and, in the case of the mid-IR source, with different measurement apparatus configurations. 
\subsection{Simulation parameters}
We briefly introduce the simulation parameters of the QKD setup.  
The transmitter randomly (50\%) prepares time-bin encoded quantum states in either the Z or X basis, with with typical employed intensities $\mu_1=0.5$ (signal) or $\mu_2=0.25$ (decoy), and sends them through the quantum channel.
At the receiver side, before measurements, we simulate a 50:50 beam splitter to implement the passive basis choice.
The receiver setup is assumed to introduce \SI{1}{\deci\bel} of losses in the Z-basis and \SI{3}{\deci\bel} in the X-basis, the latter typical of the interferometer used for relative phase measurement.\\
The detectors are modeled as SNSPD with a dead time of \SI{25}{\nano\s}. 
The efficiency and dark count rate of the SNSPD operating in the near-IR wavelengths, i.e., $\lambda = 1557.7$ nm and $\lambda = 800$ nm, are set to $\eta_{\text{det}}=0.90$ and $CR_d=1$\SI{}{\hertz}, respectively.  
\\{
For the mid-IR
wavelength source $\lambda = 3998.6$~nm, we study three different measurement setups.
The first one is a realistic {SNSPD} detector operating in the mid-IR wavelength, with an efficiency of $\eta_{\text{det}}=0.05$ and a dark count rate of {$CR_d=100$~\SI{}{\hertz}}. 
Then, the second and the third detection setups are simulations of optimized configurations, meant to be targets for further developments in mid-IR technology. In detail, 
firstly, an up-conversion process is considered, shifting the wavelength from mid- to near-IR and
therefore allowing the use of detectors operating in
the near-IR.
We take into account an efficiency of up-conversion and optical coupling of $\eta_{\text{upc}}=0.14$, resulting in an overall quantum efficiency of $\eta_{\text{det}}=0.127$ 
\cite{liu2024highly, footnote4}.
It is worth noting that the up-conversion process adds a significant background noise, heavily affecting the detector dark count rate \cite{temporao2006mid, liu2024highly}. However, these could be mitigated, in principle, via the application of adequate spectral filters and optimization of the up-conversion process. In these simulations we target an optimized up-conversion setup and we set $CR_d=1$\SI{}{\hertz} to have a direct comparison with the near-IR measurement setup. 
Secondly and finally, we simulate an optimized mid-IR SNSPD detector with an efficiency of $\eta_{\text{det}}=0.80$ \cite{chang2022efficient} and a dark count rate of {$CR_d=100$~\SI{}{\hertz}}.
}
Both the transmitter and the receiver include telescopes with an aperture diameter of $50$ cm.

\subsection{Analysis effects on the QKD link}
\label{subsec:analysis}
To study the weather effects on the secret key rates, we consider three use-case scenarios: clear weather (\SI{40}{\kilo\meter} of visibility), rain with a rainfall rate of $R = \SI{2.5}{mm/\hour}$ (associated to a visibility of \SI{6}{\kilo\meter}) and fog where the visibility is reduced to \SI{1}{\kilo\meter}. 
In Fig.\,\ref{fig:skr}, we compare the secret key rates per pulse generated by near-IR and mid-IR FSQC setup configurations, and display the results. In detail, the left column shows the performance under the different weather conditions without turbulence, while the right column includes the effects of turbulence. \\
While in general the mid-IR radiation experiences significantly lower channel losses compared to near-IR ones (see Fig. \ref{fig:closs} in Supplementary material \ref{appx:Suppl}), the figure of merit of the $skr$ demonstrates a non-trivial behaviour, due to the impact of the measurement setup under consideration, i.e., the detection system.
For short distances, setups with high-efficiency detectors -- primarily near-IR setups -- achieve the highest secure key rate. For longer distances in general, the mid-IR setup with optimized detector (cyan trace) is the best option, because it exploits both high detection efficiency and low channel losses of the mid-IR source. However, at the longest achievable distance, the scenario changes again, and the optimized mid-to-near-IR up-converted setup or the telecom one (blue and red), depending on the tested scenario, are capable of reaching a higher distance, due to the lower dark count rate detectors, as described in detail in the next paragraph.\\ 
In particular, in terms of achievable distances, the FSQC system composed of mid-IR source and optimized up-conversion process (blue trace in Fig. \ref{fig:skr}) outperforms the near-IR setups, in clear weather and low visibility (fog) scenarios, enabling longer achievable QKD links.
Indeed, in this specific case, the lower free-space channel losses of the mid-IR source compensate for the reduced quantum efficiency due to up-conversion.\\

\paragraph{Achievable distances.}
First, we notice a substantial difference in the performance of the two near-IR wavelengths: for the same detector specifications, $\lambda = 1557.6$ nm (telecom) always outperforms $\lambda = 800$ nm, since telecom shows less channel losses. 
Second, for the mid-IR source, the optimized up-conversion process, allowing the use of near-IR detectors, is the best choice for the measurement setup in terms of reachable distances for QKD. On one hand, this is because of the higher efficiency compared to the mid-IR available measurement technology (violet trace in Fig. \ref{fig:skr}) and, on the other hand, it is due to the lower dark count rate if simulating a possible highly efficient detector operating in the mid-IR (cyan trace in Fig. \ref{fig:skr}).
We therefore focus on the comparison between the telecom system and the mid-IR configuration with the optimized up-conversion at the receiver, which are the red and blue traces in Fig. \ref{fig:skr}, respectively.\\
We observe that under clear weather conditions without turbulence (Fig. \ref{fig:skr}(a)), secure quantum communication is feasible over large distances -- up to $d = \SI{500}{\kilo\meter}$ for telecom (red trace)  and $d = \SI{800}{\kilo\meter}$ for mid-IR (blue trace). However, in a more realistic scenario that accounts for turbulence (Fig. \ref{fig:skr}(b)), the QKD link is notably reduced to the order of one hundred kilometers: approximately $d = \SI{100}{\kilo\meter}$ for telecom (red trace) and $d = \SI{120}{\kilo\meter}$ for mid-IR (blue trace). 
Distances are further significantly reduced when the effects of rain and fog -- whether or not combined with turbulence -- are considered.
In the rain scenario (Figs. \ref{fig:skr}(c) and (d)), with rainfall rate  $R=2.5$ \SI{}{\milli\meter/\hour} and 6 km of visibility, the additional wavelength-independent losses, coming from the rain attenuation model, reduces the mid-IR up-converted setup advantage (blue trace). However, despite the reduced advantages, we would like to highlight that the mid-IR system with optimized detector (cyan trace) remains the best option up to distances lower then 25 km in case of no turbulence (Fig. \ref{fig:skr}(c)) and of 20 km in case of turbulence (Fig. \ref{fig:skr}(d)).
The two wavelengths show comparable performances -- allowing for a link around 25 km without turbulence, and 20 km with turbulence -- with the mid-IR system still proving almost as effective as the telecom one. 
Notice that a rainfall rate increased up to {around 10} \SI{}{\milli\meter/\hour}, associated with a visibility of \SI{2}{\kilo\meter}, would reduce the QKD link around \SI{10}{\kilo\meter}.
When the visibility is 1 km, i.e., the fog scenario, the QKD link only tolerates a distance of 10 km (7.5 km) for the mid-IR (telecom) setup. In these cases (Figs. \ref{fig:skr}(e) and (f)) with such a low visibility, turbulence does not significantly affect the QKD link length.
Since 10 km can be considered as the minimum meaningful urban distance, increasing the rainfall rate above \SI{10}{\milli\meter/\hour} or further reducing the 
visibility below 1 km would generate very high losses (> \SI{50}{\deci\bel}, see Fig. \ref{fig:closs} in Supplementary material \ref{appx:Suppl}) for both wavelengths, precluding the distribution of a quantum key for a significant urban link length.  \\

\paragraph{Mid-IR setups.}
{The results achieved by simulating the optimized up-conversion process and a QKD system operating in the mid-IR are very promising. 
These setups have} the potential to achieve significantly longer transmission distances and higher $skr$ than the telecom setup, underscoring its promise for FSQC. Realizing this advantage requires single-photon mid-IR detectors with performance comparable to their near-IR counterparts, a technological gap that we anticipate will close with continued advancements in this field.

\section{Conclusions and Perspectives}
\label{sec:Conclusions}
{One of the key objectives of this work is to explore and demonstrate the potential of the mid-IR source employment in a free-space QKD link.  
In particular, mid-IR source offers the most significant advantages in terms of channel loss, e.g. resiliency against adverse weather conditions, making it a convincing candidate for future QKD implementations.
Unlike the technologies available for more commonly used wavelengths, current mid-IR measurement technologies are not yet mature enough to fully support the employment of this wavelength. However, given this gap, this work aims to set a technological target and to stimulate further development of components for the optimized up-conversion process and for SNSPD operating in mid-IR, ultimately enabling full exploitation of the potential offered by mid-IR source in QKD systems.
}\\
As a consequence, we foresee the integration of mid-IR and near-IR radiation within a multi-wavelength FSQC framework. Such integration would aim not only to extend the achievable communication distance under a wide range of weather conditions, but also to enhance the overall robustness and adaptability of the quantum network.
Indeed, the effectiveness of an FSQC link cannot be evaluated solely based on the maximum reachable distance. An equally critical aspect is the system’s interoperability — that is, its capacity to interconnect heterogeneous quantum and classical platforms operating at different wavelengths, potentially across different technologies and use cases. A multi-wavelength architecture would provide a flexible backbone exploiting the specific advantages of each spectral region, such as the low-loss propagation of mid-IR under clear and foggy conditions and the high detection efficiency available at near-IR wavelengths.\\
By dynamically switching between or simultaneously employing different spectral channels, the FSQC network could adapt in real-time to changing environmental conditions, thereby maintaining high secret key rates and ensuring communication continuity. This vision aligns with the long-term goal of building scalable, weather-resilient, and wavelength-diverse quantum networks, capable of supporting both metropolitan and intercity links, as well as hybrid quantum systems operating across distinct physical layers.

\section*{Acknowledgements}
The Authors acknowledge financial support from EU NextGenerationEU Program with the I-PHOQS Infrastructure ``Integrated infrastructure initiative in Photonic and Quantum Sciences'' [IR0000016, ID D2B8D520]; and the project SERICS -  SEcurity and RIghts in the CyberSpace [PE00000014]. This work was also partially supported by European Commission with Laserlab-Europe Project [G.A. n.871124],  MUQUABIS Project “Multiscale quantum bio-imaging and spectroscopy” [G.A. n.101070546], QUID project ``Quantum Italy Deployment'' [G.A. No 101091408]; from the Italian ESFRI Roadmap (Extreme Light Infrastructure - ELI Project); from ASI and CNR under the Joint Project ``Laboratori congiunti ASI-CNR nel settore delle Quantum Technologies (QASINO)'' (Accordo Attuativo n. 2023-47-HH.0); from the Italian Ministero dell’Università e della Ricerca (project PRIN-2022KH2KMT QUAQK). A.Z. acknowledges financial support by Consiglio Nazionale delle Ricerche (CNR) with QuONTENT project under the ``Progetti di Ricerca@CNR'' program.

\section*{Disclosure}
The authors declare no conflicts of interest.

\section*{Data Availability}

The datasets generated during and/or analysed during the current study are available from the corresponding author on reasonable request.

\bibliographystyle{ieeetr_CA}
\bibliography{references}

\begin{thebibliography}{10}
\bibitem{footnote5}
The range is declared according to ISO 20473.
\bibitem{footnote1}
The atmospheric attenuation due to Rayleigh scattering at 500 nm (visible) is 1000 times the one at 4.0 $\mu$m (mid-IR). The attenuation at 1.55 $\mu$m (telecom) is 200 times larger than at 4.0 $\mu$m.
\bibitem{footnote4}
Commercially available mid-to near-infraded wavelength converter modules are prensent on the market, such as the the TUNE produced by nlir with a declared conversion effiency of 3 \%.
\end{thebibliography}

\appendix
\section{Supplementary material}
\label{appx:Suppl}
\paragraph{Geometrical loss computations.}
\label{appx:geom}
The radius of a Gaussian beam propagating in free-space depends on the wavelength and the distance between transmitter and receiver. Shifted by a factor $Z$, the Gaussian beam radius function is given by: 
\begin{align*}
    w(z) & = w_0 \left[1+\left(\frac{z-Z}{Z}\right)^2\right],
\end{align*}
with $Z$ the Rayleigh distance and $w_0$ the beam waist.
The ratio between the beam surface and the receiver surfaces, given $r_{Rx} = \sqrt{2} w_0$, is therefore: 
\begin{align*}
    L_z & = \frac{S_d}{S_{Rx}}  =\frac{\pi \ w(z)^2}{\pi \ (r_{Rx})^2} =  \frac{w_0^2}{2 w_0^2} \left[1+\left(\frac{z-Z}{Z}\right)^2\right] \\
    & = \frac{1}{2} \left[1+\left(\frac{z-Z}{Z}\right)^2\right], 
\end{align*}
and the Rayleigh distance can be written as: 
\begin{equation*}
    Z = \frac{\pi \ w_0^2}{\lambda}=\frac{\pi \ (r_{Rx}/\sqrt{2})^2}{\lambda}=\frac{\pi \ (r_{Rx})^2}{2 \ \lambda}.
\end{equation*}
\paragraph{Channel attenuation comparison.}
For the sake of completeness, we show here the comparison of channel losses, for near-IR and mid-IR sources, in weather conditions corresponding to the results given in Fig. \ref{fig:skr}. 
\begin{figure}[H]
    \centering
    \subfigure[]{\includegraphics[width=0.23\textwidth]{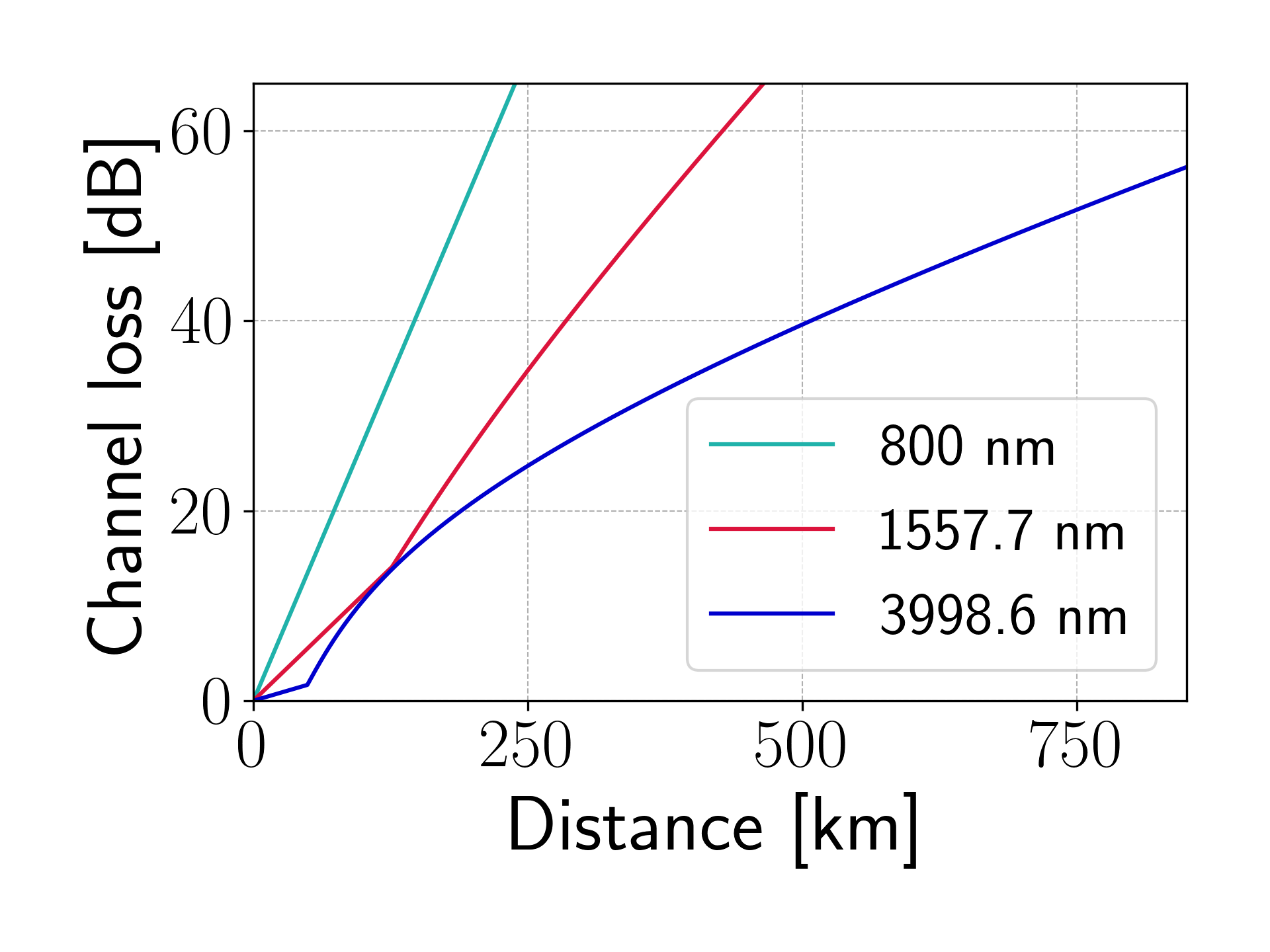}}
    \subfigure[]{\includegraphics[width=0.23\textwidth]{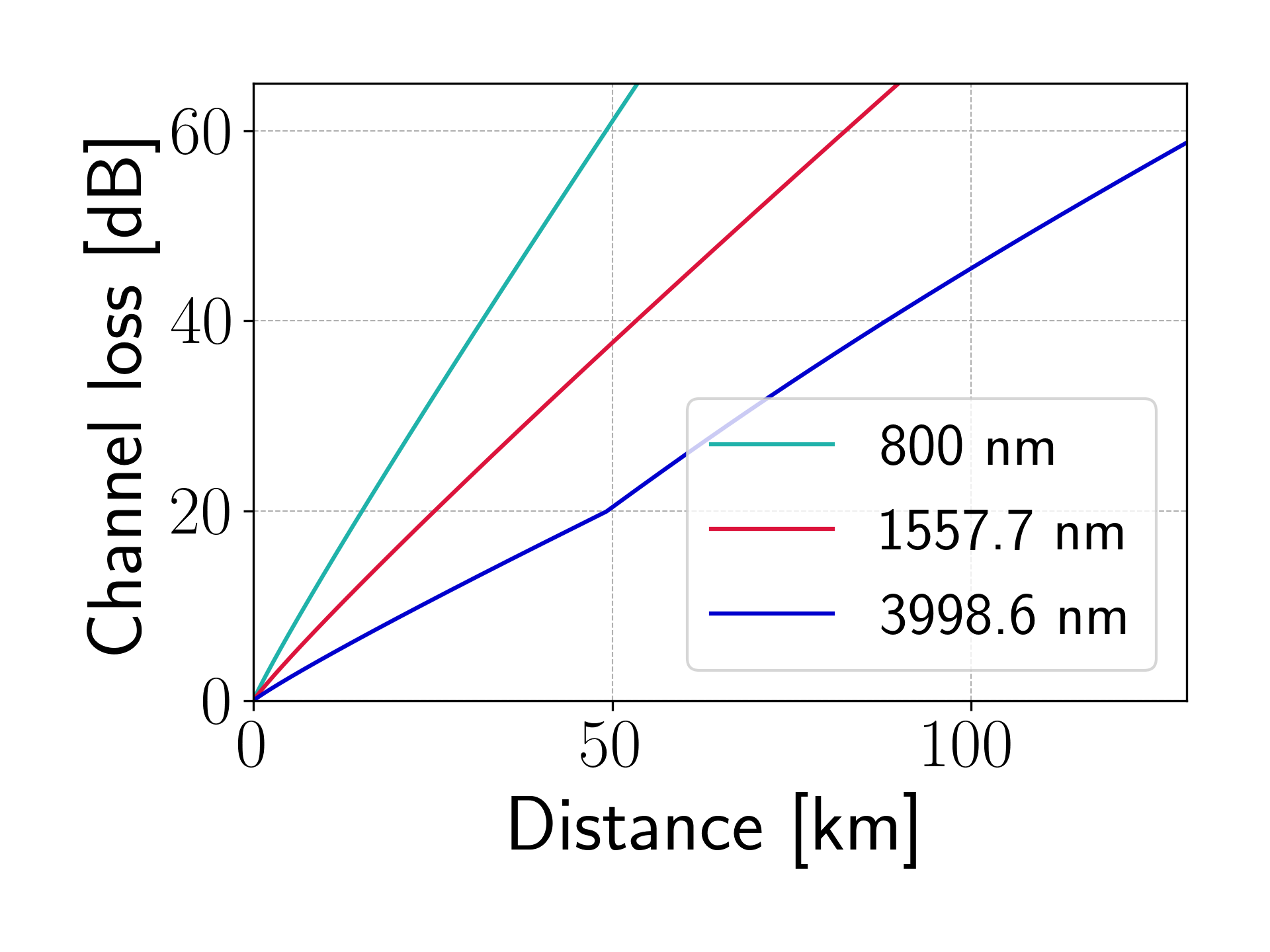}}
    
    \subfigure[]{\includegraphics[width=0.23\textwidth]{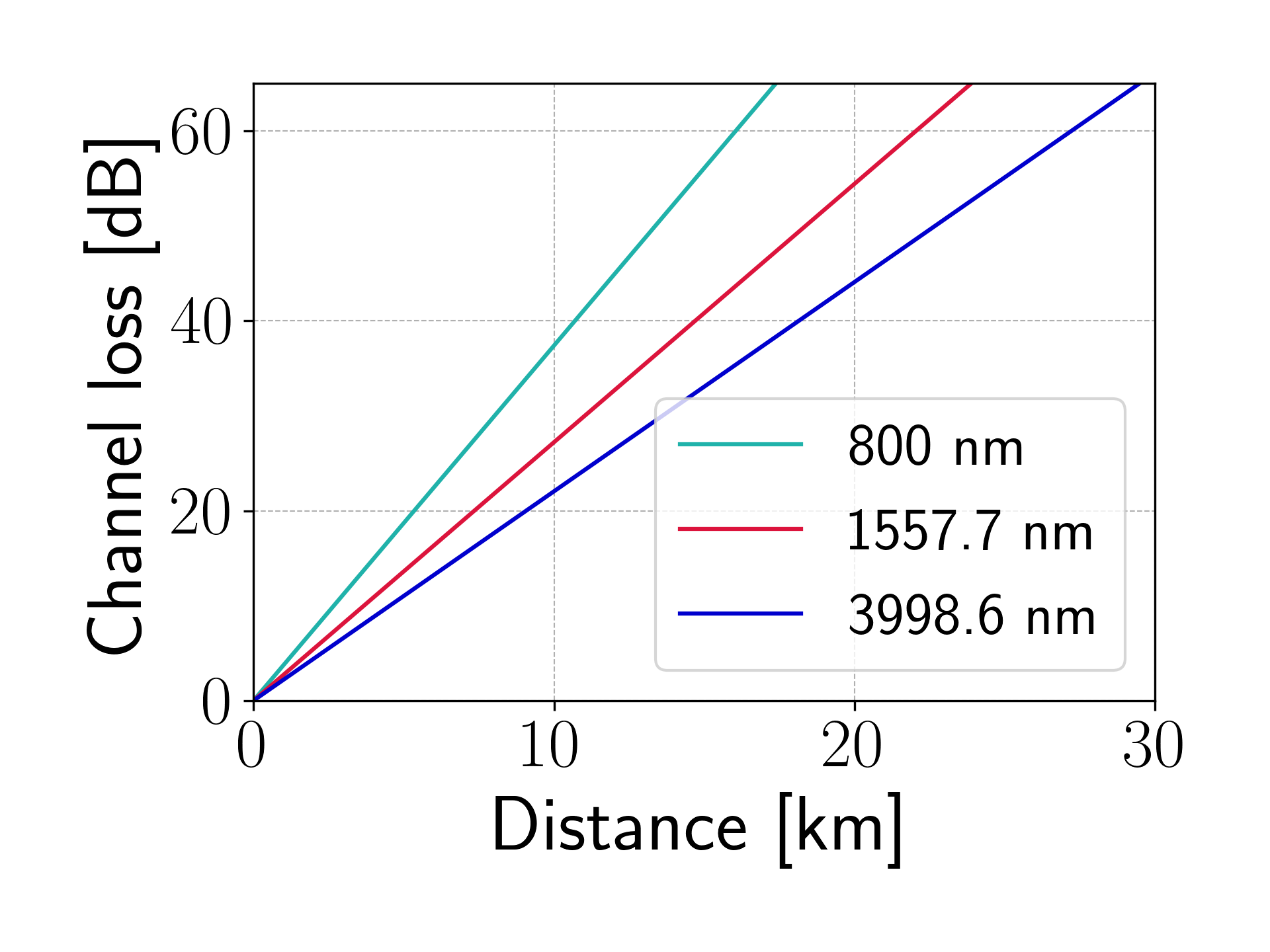}}
    \subfigure[]{\includegraphics[width=0.23\textwidth]{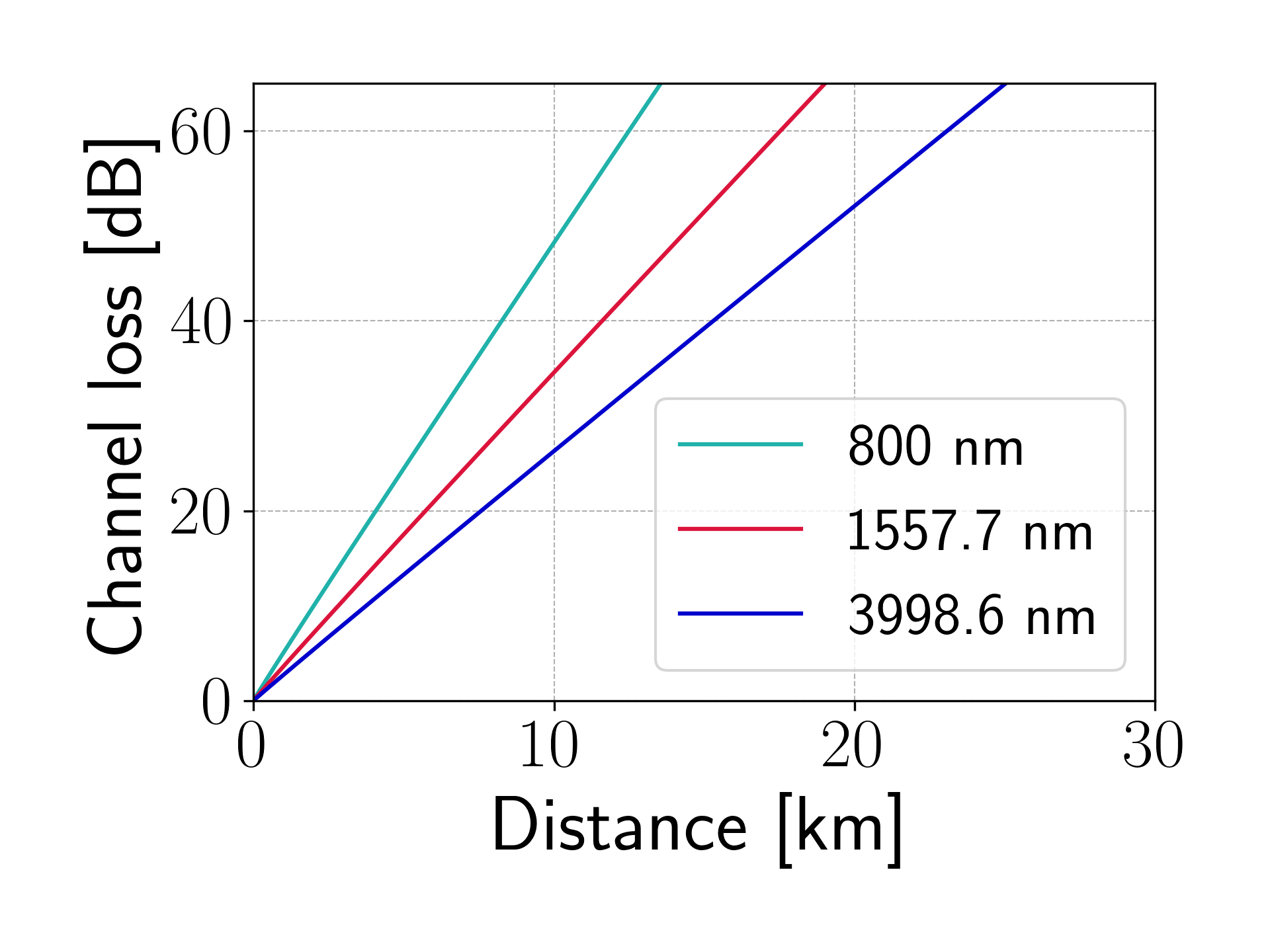}}
    
    \subfigure[]{\includegraphics[width=0.23\textwidth]{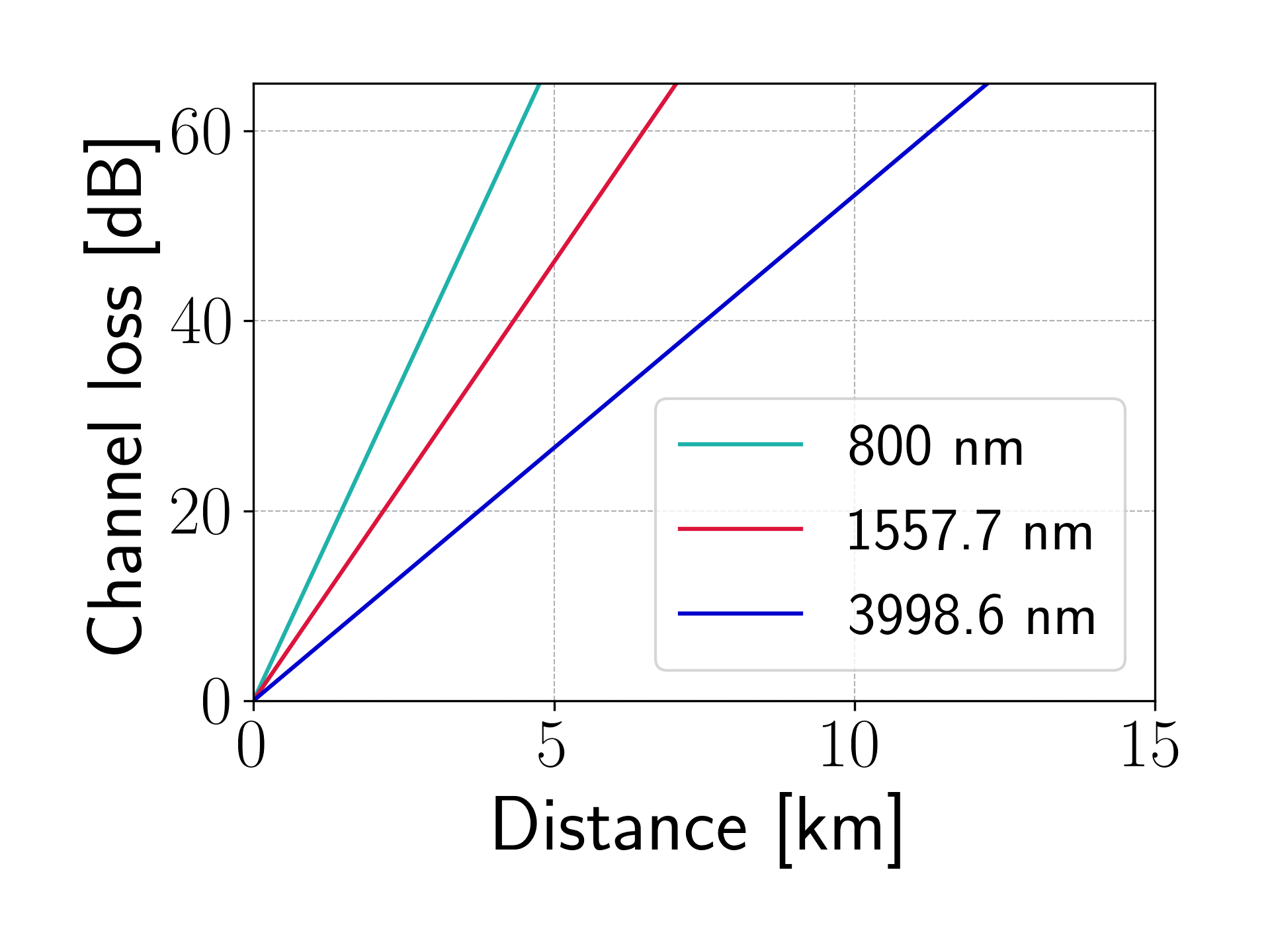}}
    \subfigure[]{\includegraphics[width=0.23\textwidth]{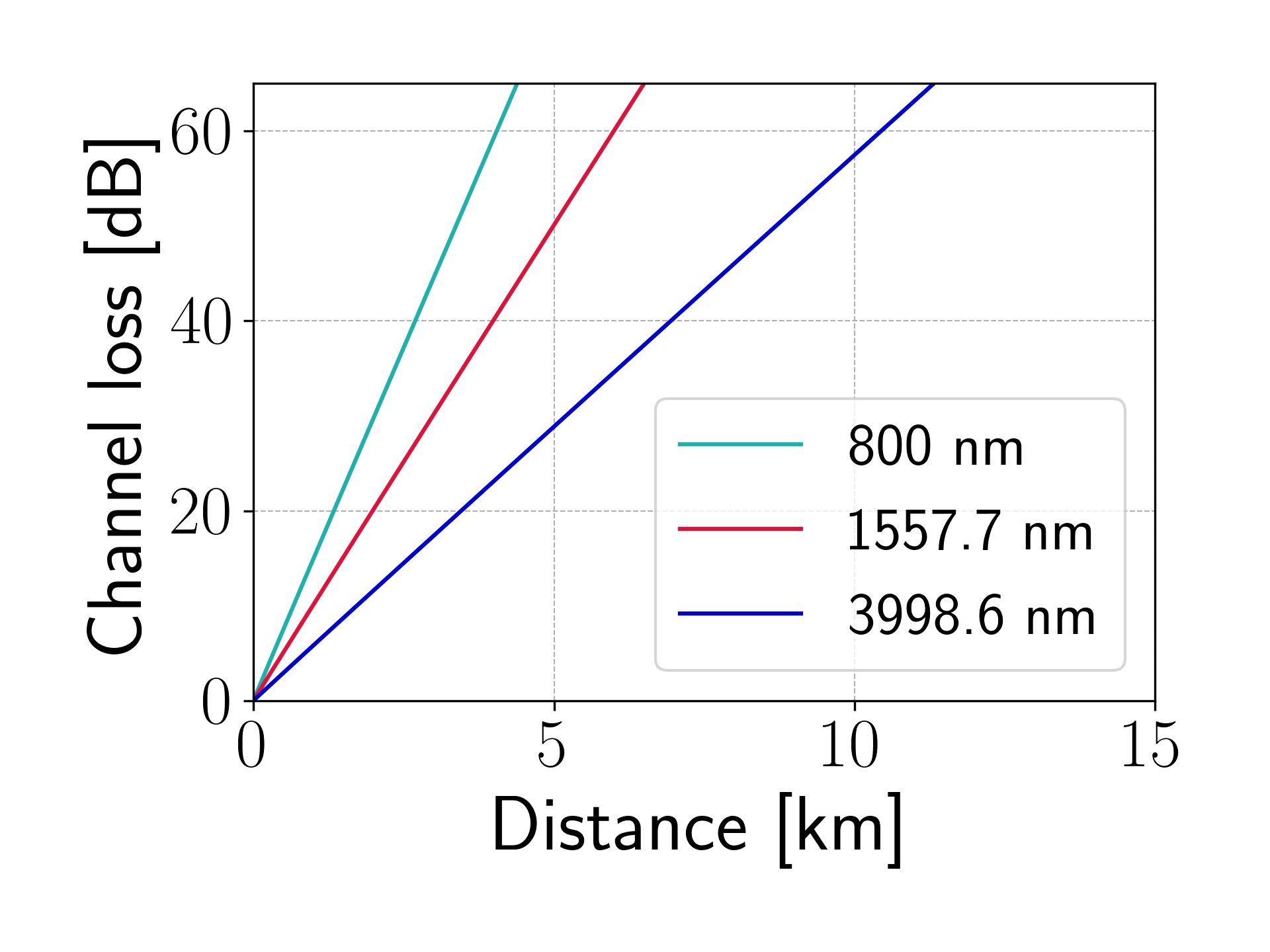}}
    \caption{Comparison of channel loss in function of the link length, for the near-IR wavelengths  $\lambda = \SI{1557.7}{nm}$ and $\lambda =\SI{800}{nm}$, and mid-IR $\lambda =\SI{3998.6}{nm}$ in the weather conditions: (a) clear weather (V = \SI{40}{\kilo\meter} visibility); (b) clear weather with turbulence; (c) rain (R = \SI{2.5}{\milli\meter/\hour} of rainfall rate corresponding to a visibility of \SI{6}{\kilo\meter}); (d) rain with turbulence; (e) fog (V = \SI{1}{\kilo\meter} visibility); (f) fog with turbulence.
    }
    \label{fig:closs}
\end{figure}

\end{document}